\documentclass[superscriptaddress,amsmath,amssymb,aps,prl,twocolumn]{revtex4-1}

\usepackage{times}
\usepackage{graphicx}
\usepackage{dcolumn}

\begin{document}

\title{Decoupling optical function and geometrical form using conformal flexible dielectric metasurfaces}

\author{Seyedeh Mahsa Kamali}
\affiliation{T. J. Watson Laboratory of Applied Physics and Kavli Nanoscience Institute, California Institute of Technology, 1200 E California Blvd., Pasadena, CA 91125, USA}
\author{Amir Arbabi}
\affiliation{T. J. Watson Laboratory of Applied Physics and Kavli Nanoscience Institute, California Institute of Technology, 1200 E California Blvd., Pasadena, CA 91125, USA}
\author{Ehsan Arbabi}
\affiliation{T. J. Watson Laboratory of Applied Physics and Kavli Nanoscience Institute, California Institute of Technology, 1200 E California Blvd., Pasadena, CA 91125, USA}
\author{Yu Horie}
\affiliation{T. J. Watson Laboratory of Applied Physics and Kavli Nanoscience Institute, California Institute of Technology, 1200 E California Blvd., Pasadena, CA 91125, USA}
\author{Andrei Faraon}
\email{Corresponding author: A.F: faraon@caltech.edu}
\affiliation{T. J. Watson Laboratory of Applied Physics and Kavli Nanoscience Institute, California Institute of Technology, 1200 E California Blvd., Pasadena, CA 91125, USA}

\begin{abstract}
Physical geometry and optical properties of objects are correlated: cylinders focus light to a line, spheres to a point, and arbitrarily shaped objects introduce optical aberrations. Multi-functional components with decoupled geometrical form and optical function are needed when specific optical functionalities must be provided while the shapes are dictated by other considerations like ergonomics, aerodynamics, or esthetics. Here we demonstrate an approach for decoupling optical properties of objects from their physical shape using thin and flexible dielectric metasurfaces which conform to objects' surface and change their optical properties. The conformal metasurfaces are composed of silicon nano-posts embedded in a polymer substrate that locally modify near-infrared ($\it{\lambda}$ = 915 nm) optical wavefronts. As proof of concept, we show that cylindrical lenses covered with metasurfaces can be transformed to function as aspherical lenses focusing light to a point. The conformal metasurface concept is highly versatile for developing arbitrarily shaped multi-functional optical devices.
\end{abstract}

\maketitle
The correlation between the geometry of an object and its optical functionality~\cite{born1999principles} has introduced long-standing design challenges to optical engineers developing multi-functional components~\cite{thompson2012freeform}. The traditional solution has been to compromise and optimize the component material and geometry by considering all the physical requirements. This was originally studied in the context of conformal and freeform optics where optical components with non-standard surfaces were developed for integration of optics into flying objects with specific aerodynamic shapes~\cite{ shannonSPIE1999,KanppSPIE2002}. More recently, this issue has attracted new attention due to its application in integration of optics into various consumer electronic products and medical equipment with stringent packaging and design requirements. Furthermore, controlling optical properties of objects without physically modifying them can enable the visual blending of an object with its background~\cite{fan2012invisible,ni2015ultrathin,valentine2009optical,ergin2010three} or changing its appearance through generation of a holographic virtual image \cite{zheng2015metasurface,teo2015controlling}. In the context of conformal optics, the conventional solution is to stack several bulky optical elements with non-standard surface profiles underneath the outermost surface of the object~\cite{KanppSPIE2002}. Such solutions usually have challenging fabrication processes requiring custom-made fabrication equipment, are bulky, and do not provide a unified and versatile approach that can be applied to arbitrary geometries. Conformal metasurface approach can provide a solution for decoupling the geometric shape and optical characteristics of arbitrary objects.

Metasurfaces are two dimensional arrays of scatterers rationally designed to locally modify phase and polarization of electromagnetic waves~\cite{kildishev2013planar,yu2014flat,Lin2014a,arbabi2015dielectric}. They enable wafer-scale production of lithographically-defined thin diffractive optical elements using conventional nano-manufacturing techniques. These manufacturing techniques are optimized for patterning flat substrates and are not applicable for direct fabrication of metasurfaces on non-planar structures required for conformal optics. However, the two dimensional nature and the minute thickness of optical metasurfaces make them suitable for transferring to non-planar substrates. Several different plasmonic and dielectric metasurface platforms for optical wavefront manipulation have been recently proposed~\cite{kildishev2013planar,yu2014flat,karimi2014generating,Fattal2010,Lin2014a,arbabi2015subwavelength,arbabi2015dielectric,Vo2014}. Among different platforms, dielectric metasurfaces based on high contrast transmitarrays are highly versatile~\cite{arbabi2015subwavelength,arbabi2015dielectric,Vo2014} as they provide simultaneous manipulation of phase and polarization of light with high efficiencies, and can sample optical wavefronts with sub-wavelength spatial resolution~\cite{arbabi2015dielectric}. In these metasurfaces, each meta-atom is a high index nano-post acting as a short waveguide, which locally imposes a certain phase shift and polarization rotation. Several efforts have been made to transfer metasurfaces (mostly plasmonic ones) to flexible substrates with the aim of tuning their frequency response using substrate deformation \cite{di2010flexible,pryce2010highly,xu2011flexible,walia2015flexible,zhu2015flexible,gao2015optics, gutruf2015mechanically}. Plasmonic metasurfaces, however, have low efficiencies especially in transmission mode, which in many situations make them impractical.

Here, we introduce flexible metasurfaces based on a dielectric high contrast transmitarray platform that can be conformed to a non-planar arbitrarily shaped object to modify its optical properties at will. We present a general design procedure and a high yield fabrication process for the conformal flexible metasurface platform. As proof of principle, we experimentally demonstrate flexible metasurfaces that wrap over cylindrical surfaces and convert them to aspherical lenses.

\begin{figure*}[htp]
\centering
\includegraphics[width=1.8\columnwidth]{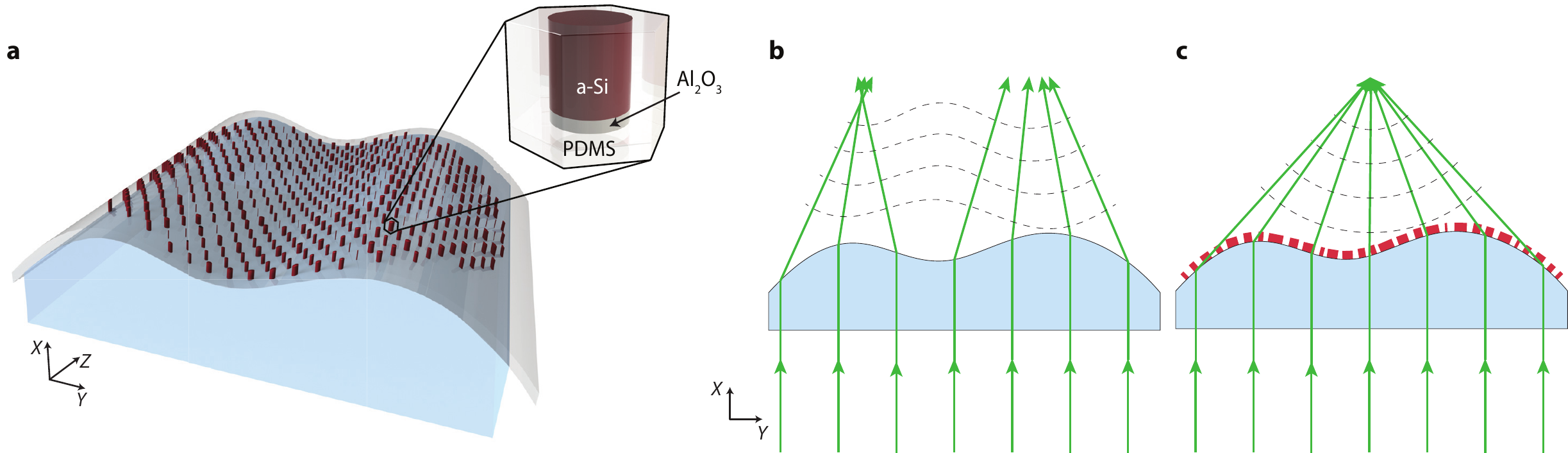}
\caption{\textbf{Conformal optics with optical dielectric metasurfaces.} \textbf{a}, A schematic illustration of a dielectric metasurface layer conformed to the surface of a transparent object with arbitrary geometry. (Inset) The building block of the metasurface structure: an amorphous silicon (a-Si) nano-post on a thin layer of aluminum oxide ($Al_2O_3$) embedded in a low index flexible substrate (polydimethylsiloxane (PDMS) for instance).  \textbf{b}, Side view of the arbitrarily shaped object showing how the object refracts light according to its geometry and generates an undesirable wavefront. \textbf{c}, The same object with a thin dielectric metasurface layer conformed to its surface to change its optical response to a desired one.}
\label{fig:Fig1_Concept}
\end{figure*}

Figure \ref{fig:Fig1_Concept}a shows a schematic illustration of a non-planar arbitrarily shaped transparent object wrapped by a flexible metasurface based on this platform. The metasurface layer is composed of an array of dissimilar cylindrical amorphous silicon (a-Si) nano-posts with different diameters placed on a subwavelength periodic hexagonal lattice, and embedded in polydimethylsiloxane (PDMS) as a flexible substrate (Fig.  \ref{fig:Fig1_Concept}a, inset). The arbitrary shape of the object's surface distorts the wavefront of the transmitted light in an undesirable way (Fig.  \ref{fig:Fig1_Concept}b). By conforming the metasurface onto the object's outermost surface, the distortion can be compensated and the wavefront of the transmitted light can be shaped to a desired form, similar to phase compensating antenna arrays employed in the microwave regime~\cite{josefsson2006conformal}. For example, the metasurface can be designed to correct the distortions introduced by the arbitrarily shaped object and make it act similar to an aspherical lens that focuses light to a point as schematically shown in Fig.  \ref{fig:Fig1_Concept}c. 

The desired phase profile of the conformal metasurface is found with the knowledge of the geometry of the transparent object over which it is wrapped, and the desired optical response. First, the object without the metasurface is considered, and the phase profile of the optical waves transmitted through the object is computed along the surface of the object. For objects with dimensions significantly larger than the optical wavelengths, this phase profile can be found using ray optics approximation and by computing the optical path length and the corresponding optical path difference (OPD) of the rays passing through different points along the outermost surface of the object with respect to the chief ray.  Then, using a similar OPD-based approach, the phase profile required to achieve the desired specific functionality is obtained along the surface of the object. For example, if we want the object to focus light to a point, a converging spherical wavefront is desired, which is sampled along the arbitrary surface of the object. The metasurface layer, when wrapped on the surface of the object, should locally impose an additional optical phase shift equal to the difference between the original phase of the object and the desired phase profile. Therefore, the desired metasurface phase profile is expressed as a function of two coordinate values defining the non-planar surface of the object. To obtain the appropriate phase profile of the metasurface before its transfer to the non-planar surface, an appropriate coordinate transformation should be applied. For example, if the flexible substrate of the metasurface is under no stress after being mounted on the object's surface, then the appropriate coordinate transformation conserves length along the surface of the object.

\begin{figure*}[htp]
\centering
\includegraphics[width=2\columnwidth]{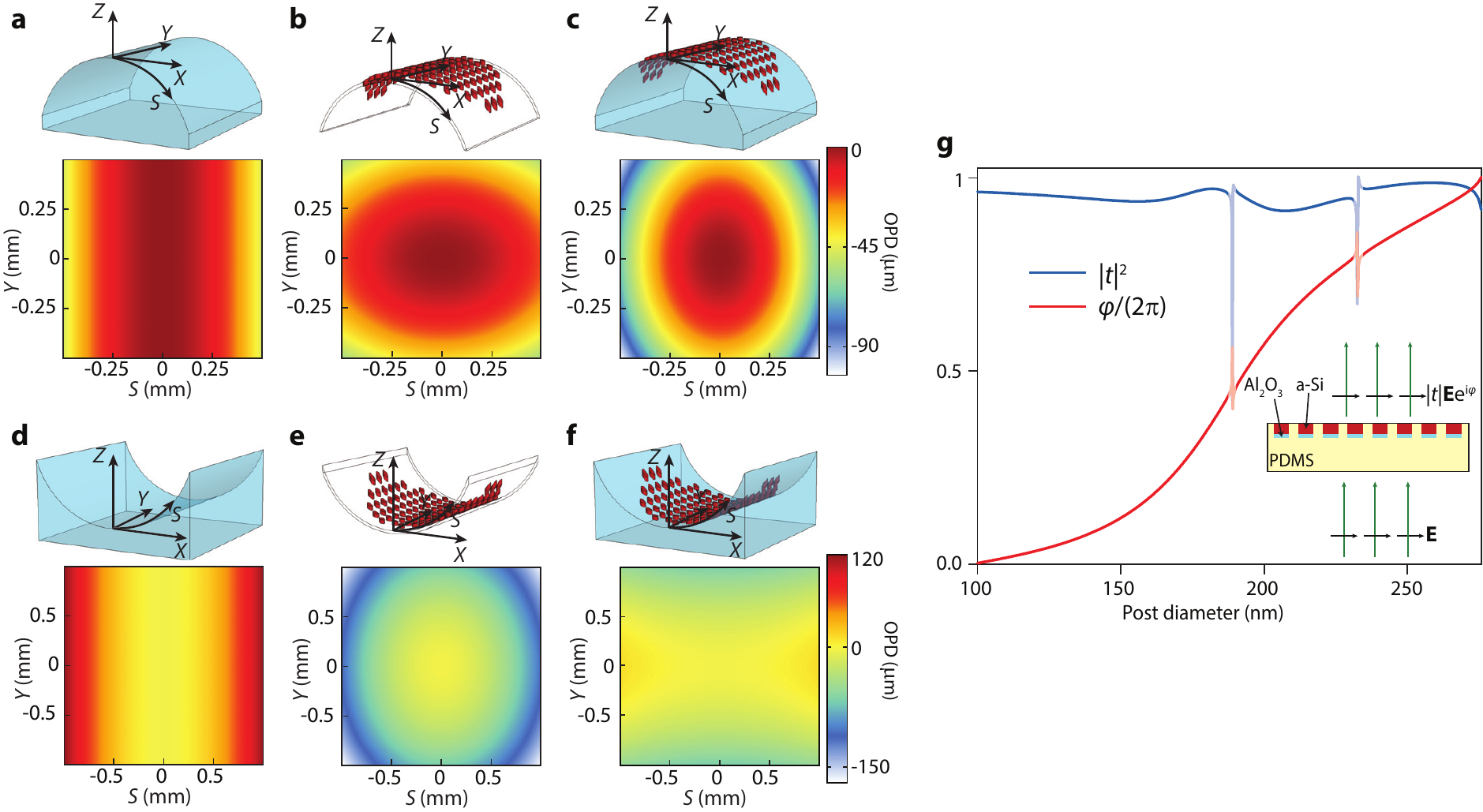}
\caption{\textbf{Design procedure of conformal metasurfaces.} \textbf{a}, The optical path difference (OPD) (in $\mu$m) of the rays passing through a converging cylindrical object. \textbf{b}, The difference OPD needed at the surface of the convex cylindrical object compensated by the conformal metasurface. \textbf{c}, Desired OPD at the surface of the object which is provided by the object and conformal metasurface combination. \textbf{d}, \textbf{e}, and \textbf{f} show similar plots for a diverging cylinder. $\it{``S"}$ is the arch length on the cylinder surface in a plane perpendicular to the $\it{Y}$ axis. \textbf{g}, Simulated intensity transmission and phase of the transmission coefficient for a periodic array of amorphous silicon (a-Si) nano-posts embedded in polydimethylsiloxane (PDMS) as shown in the inset. The nano-posts are composed of 720 nm a-Si on 100 nm aluminum oxide ($\mathrm{Al_2O_3}$), and are arranged in a hexagonal lattice. The simulation wavelength is 915 nm. This graph is used to relate the phase shift values (and the respective OPDs) needed at different points on the conformal metasurface to the nano-post diameters. See Methods for simulation details.}
\label{fig:Fig2_Design}
\end{figure*}

Using this design procedure, we computed two sets of conformal metasurface phase profiles for both a convex and a concave cylindrical glass. The metasurfaces modify the wavefronts of the cylindrical objects to make them behave as aspherical lenses. Figure \ref{fig:Fig2_Design}a (\ref{fig:Fig2_Design}d) shows the OPD of the rays passing through the convex (concave) cylinder at its top surface. Considering the desired converging spherical wavefront, the desired OPD of the rays at the surface of the convex (concave) cylinder is calculated and shown in Fig. \ref{fig:Fig2_Design}c (\ref{fig:Fig2_Design}f). The difference between the OPDs of the convex (concave) cylindrical object and the converging spherical phase profile is shown in Fig. \ref{fig:Fig2_Design}b (\ref{fig:Fig2_Design}e). The conformal metasurfaces should impose phase shifts equivalent to these OPDs at the operation wavelength (see methods for simulation details). Since the cylindrical surfaces are isometric with a plane, the metasurfaces can be mounted on them under negligible stress. Therefore, only a simple geometric transformation ($\it{XY}$ to $\it{SY}$ in Fig. \ref{fig:Fig2_Design}a) is used to map the coordinates on a cylinder surface to a plane.

The optical coupling among the nano-posts is weak in the high contrast transmitarray metasurface platform, and each nano-post scatters light almost independent of its neighboring nano-posts. The weak coupling is due to the high index contrast between the nano-posts and their surroundings, and it is manifested in the localization of the optical energy inside the nano-posts and the weak dependence of the transmission of the nano-post arrays to their spacing (i.e. lattice constant) as has been previously discussed \cite{arbabi2015subwavelength} in more detail. This simplifies the design by allowing to directly relate the local transmission coefficient to the diameter of the nano-post at each unit cell of the metasurface. Figure \ref{fig:Fig2_Design}g shows the simulated intensity transmission coefficient and phase of the transmission coefficient for periodic arrays of 720 nm tall nano-posts embedded in PDMS with diameters ranging from 100 nm to 275 nm (see methods for simulation details). The nano-posts are arranged in a hexagonal lattice with 550 nm lattice constant, and the simulation wavelength is 915 nm. Refractive indices of a-Si and PDMS are 3.56 and 1.41 at the simulation wavelength, respectively. The whole 0 to 2$\pi$ phase range can be covered by changing the nano-post diameters while keeping the intensity transmission coefficient above 91\%. These results are obtained assuming normal incidence.

To get more insight into the operation mechanism, each nano-post can be considered as a truncated circular cross-section waveguide \cite{lalanne1999waveguiding}. Because of the truncation of both ends, the nano-post supports multiple low quality factor Fabry-Perot resonances which interfere and lead to high transmission of the nano-post array (see Supplementary Note 1 and  Supplementary Fig. 1). We also note that in contrast to Huygens' metasurfaces where only two resonant modes are employed (one with a significant electric dipole and one with significant magnetic dipole) \cite{decker2015high}, the resonant modes of the nano-posts contain dipole, quadrupole and higher order electric and magnetic multipoles in their multipole expansion. Although the modal expansion approach provides some intuitive understanding of the operation principle, it does not offer guidelines for designing of the nano-post arrays. Moreover, an effective medium method does not capture the underlying physics of the periodic structures that support more than one propagating mode \cite{Vo2014, lalanne1999waveguiding, lalanne1998blazed}. Therefore it is not applicable to most of the nano-posts widths we used in designing the metasurface, as a periodic array of nano-posts with diameters greater than 180 nm would be multimode. Considering these, and the limited number of design parameters (i.e. nano-post height and the lattice constant), we prefer the direct approach of finding the transmission of the nano-post arrays (as shown in Fig. \ref{fig:Fig2_Design}g) over the modal expansion technique.

\begin{figure*}[htp]
\centering
\includegraphics[width=1.7\columnwidth]{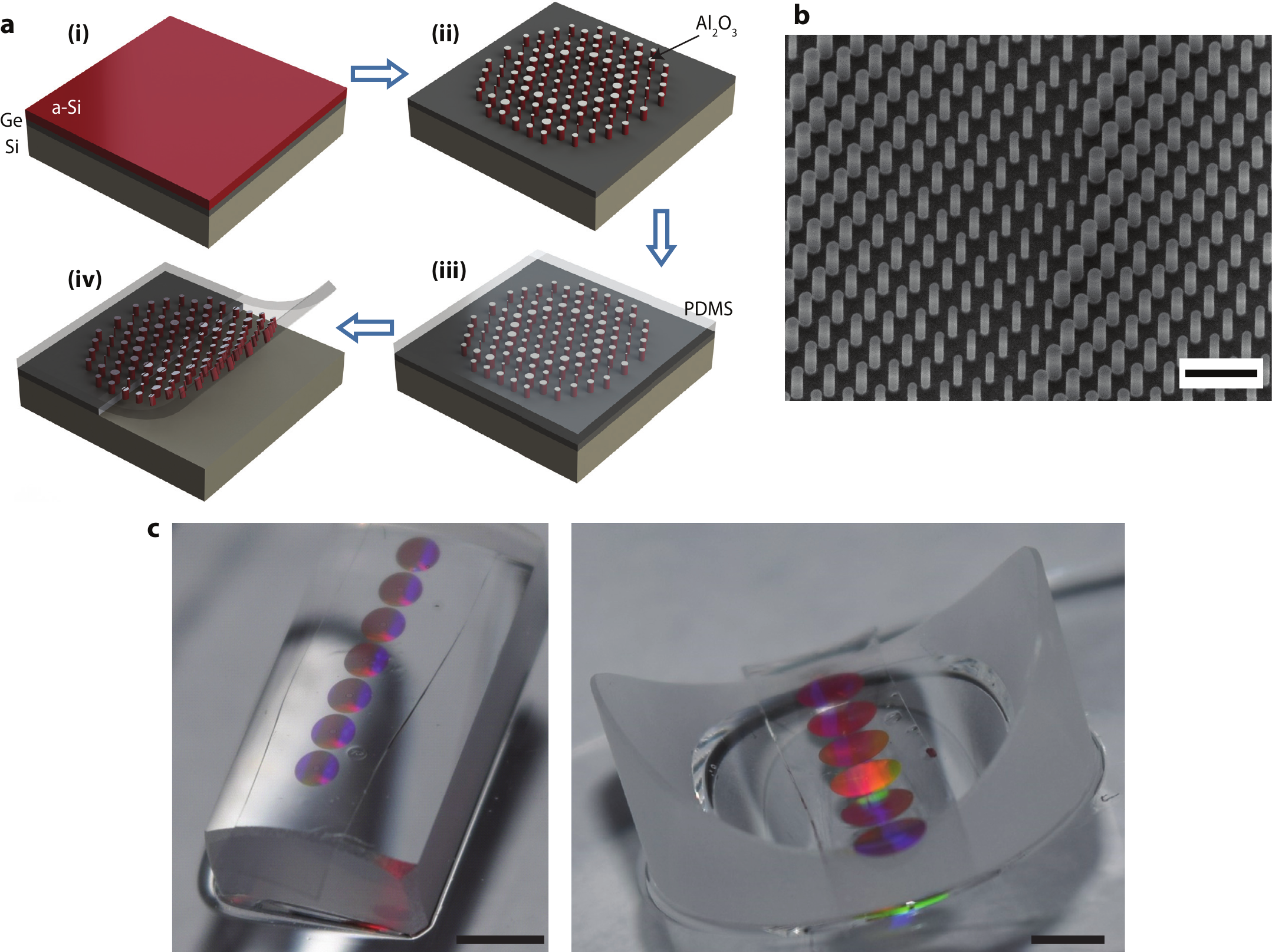}
\caption{\textbf{Overview of the fabrication process and images of the fabricated metasurface.} \textbf{a}, Steps involved in the fabrication of conformal metasurfaces: (\textbf{i}) Germanium (Ge) and amorphous silicon (a-Si) are deposited on a silicon wafer. (\textbf{ii}) a-Si nano-posts are patterned and dry etched using an aluminum oxide hard mask. (\textbf{iii}) Polydimethylsiloxane (PDMS) is spin coated on the substrate. (\textbf{iv}) The sacrificial Ge layer is dissolved to release the nano-posts which are embedded in the flexible PDMS layer. \textbf{b}, A scanning electron microscope image of the silicon nano-posts with the aluminum oxide mask before spin coating PDMS. Scale bar, 1 $\mu$m. \textbf{c}, Optical images of two flexible metasurfaces conformed to a convex glass cylinder (left) and a concave glass cylinder (right). In both cases, the metasurfaces make cylinders behave like converging aspherical lenses. Scale bars, 2 mm.}
\label{fig:Fig3_Fab}
\end{figure*}

Low sensitivity to the incident angle is a necessary property for a conformal metasurface since the incident angle would be varying across the metasurface when it is wrapped over a non-planar object. For the metasurface platform considered here, the transmission coefficient of transverse electric (TE) polarized light is weakly dependent on the incidence angle, and transmission coefficient of transverse magnetic (TM) polarized light shows some angle dependent resonances (Supplementary Note 2 and  Supplementary Fig. 2). These resonances introduce a small phase error and lower transmission, but as we experimentally show, they only slightly reduce the metasurface efficiency for TM polarization. For very steep angles, the metasurface efficiency decreases as analyzed in our previous work \cite{arbabi2015subwavelength}. The general metasurface design procedure is as follows. First, the coordinate-transformed desired metasurface phase was sampled at the lattice sites of the periodic hexagonal lattice. Then, the diameter of the nano-post at each site was obtained using the corresponding sampled phase value at that site and the phase-diameter relation shown in Fig. \ref{fig:Fig2_Design}g. To ensure a one to one relationship between the phase and nano-post diameters, and to keep the transmission high, nano-post diameters corresponding to the sharp resonances in Fig. \ref{fig:Fig2_Design}g were not used. Using this procedure, metasurfaces with phase profiles shown in Fig. \ref{fig:Fig2_Design}b and \ref{fig:Fig2_Design}e were designed to be conformed to convex and concave cylindrical objects. These metasurfaces modify the optical response of the cylinders such that they behave as aspherical lenses and focus light to single points (see Methods for the details of designed lenses and cylindrical surfaces).

\begin{figure*}[htp]
\centering
\includegraphics[width=1.9\columnwidth]{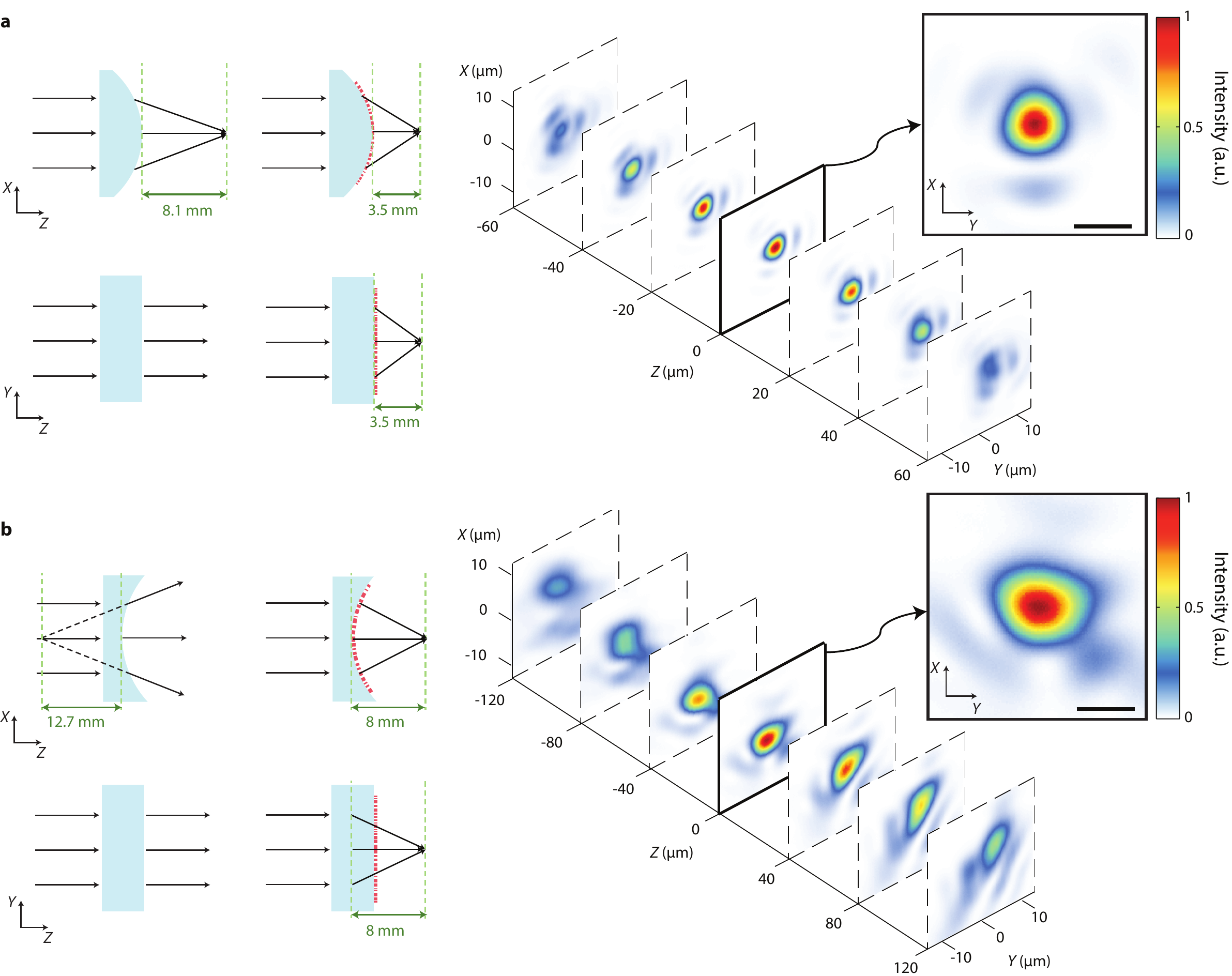}
\caption{\textbf{Measurement results of conformal dielectric metasurfaces.} \textbf{a}, A converging cylindrical lens with a radius of 4.13 mm and a focal distance of 8.1 mm is optically modified using a conformal metasurface with a diameter of 1 mm. The cylinder plus the metasurface combination behaves as an aspherical lens with a focal length of 3.5 mm. The coordinate system is the same as in Fig. 2. \textbf{b}, A different metasurface is mounted on a concave glass cylinder with a radius of 6.48 mm and a focal distance of $-$12.7 mm, which makes it focus to a spot 8 mm away from its surface (as an aspherical lens). Schematic illustrations (side and top views) are shown on the left, and intensities at planes parallel to the focal plane and at different distances from it are shown on the right. Intensities at the focal plane are depicted in the insets. Measurements are performed at the wavelength of 915 nm. For the measurement details see Methods. Scale bars, 5 $\mu$m.}
\label{fig:Fig4_Measurement}
\end{figure*}

Figure \ref{fig:Fig3_Fab}a schematically illustrates the key steps in fabricating thin, flexible, and conformable metasurfaces. A germanium sacrificial layer is deposited on a silicon wafer and then an a-Si layer is deposited over the germanium (Fig. \ref{fig:Fig3_Fab}a, (i)). The a-Si layer is patterned using electron-beam lithography followed by dry etching using an alumina hard mask (Fig. \ref{fig:Fig3_Fab}a, (ii)). The sample is subsequently spin coated with two layers of PDMS (a diluted thin layer followed by a thicker layer (Fig. \ref{fig:Fig3_Fab}a, (iii)). Then, the sample is immersed in a diluted ammonia solution which dissolves the germanium layer and releases the flexible metasurface with minimal degradation of the metasurface and the PDMS layer (Fig. \ref{fig:Fig3_Fab}a, (iv)). A scanning electron microscope image of the fabricated device before spin coating the PDMS layer is shown in Fig. \ref{fig:Fig3_Fab}b. Optical images of metasurfaces conformed to the convex and concave glass cylinders are shown in Fig. \ref{fig:Fig3_Fab}c. The whole fabrication process has a near unity yield, with almost all of the metasurfaces retaining a large majority of the nano-posts (Supplementary Note 3 and Supplementary Fig. 3). Moreover, it does not degrade the optical quality of the metasurface layer. The optical quality of the flexible metasurface layer was tested by transferring a flat metasurface lens to a flat substrate. See Supplementary Fig. 4 for the measurement results and focusing efficiency of the transferred flat metasurface lens. To demonstrate the capabilities of this platform, two different conformal metasurfaces operating at the near infrared wavelength of 915 nm were fabricated and characterized. The first 1-mm-diameter metasurface conforms to a converging cylindrical lens with a radius of 4.13 mm. The cylinder by itself focuses light to a line 8.1 mm away from its surface (Fig. \ref{fig:Fig4_Measurement}a).  The presence of the metasurface modifies the cylinder to behave as an aspherical lens focusing light to a point 3.5 mm away from the surface of the cylinder (Fig. \ref{fig:Fig4_Measurement}a). The second device is a 2-mm diameter metasurface conforming to a diverging glass cylinder with a radius of 6.48 mm and a focal length of $-$12.7 mm (Fig. \ref{fig:Fig4_Measurement}b). With the metasurface on top, the concave cylinder focuses light to a point 8 mm away from the cylinder surface (Fig. \ref{fig:Fig4_Measurement}b). 

The devices were characterized under 915 nm collimated laser beam illumination by recording intensity profiles at different planes parallel to their focal planes. Figure \ref{fig:Fig4_Measurement} also shows the measured intensity profiles. The focal plane intensity profiles are shown as insets. A tight focus is observed at the designed focal length. Focusing efficiencies of 56\% and 52\% under TE illumination (i.e. electrical field parallel to the cylinder axis) were measured for the two devices, respectively. The focusing efficiency is defined as the ratio of the power focused by the device to the incident power on the device (see Methods for the measurement details). Under TM illumination, numerical estimations based on the angular response of a uniform array shown in Supplementary Fig. 2 indicate a slight degradation of the device performance for larger angles between the metasurface and the incident beam. The devices were measured with TM input beam polarizations and, as expected, showed similar behavior as under TE illumination with focusing efficiencies of 56\% and 50\%. The difference in TE and TM polarization efficiencies increases as the incidence angle becomes steeper (Supplementary Fig. 5); the focus pattern, however, remains almost the same under both polarizations (Supplementary Fig. 6). The corresponding measured full width at half maximum (FWHM) of the focal spots are approximately 3.5 $\mu$m and 5 $\mu$m comparable to diffraction limited FWHM of 3.2 $\mu$m and 3.7 $\mu$m, respectively. Slight aberrations observed in the focal plane intensity profiles are mostly due to imperfections in the alignment of the metasurface to the non-planar substrates. Reduction of efficiency in conformal metasurfaces compared to the transferred flat metasurfaces (Supplementary Fig. 4) is mostly due to the imperfections in the alignment, slight movements of the nano-posts within the flexible substrate during the substrate handling, and the difference between the actual non-planar substrate profile and the profile assumed for design.

Although here we have used cylindrical substrates as proof of principle, this platform is not limited to surfaces that can be projected to a plane using isometric transformations. Conformal metasurfaces can be designed for other types of objects (for instance spheres where the metasurface needs to be stretched for conforming) with a similar method. High stretchability and flexibility of thin PDMS layers ($\sim$50 $\mu$m) make them suitable for conforming to non-isometric surfaces. In such cases, however, a mechanical analysis of the metasurface deformation upon mounting on the object should be carried out. The coordinate transformation that projects the conformal lattice to the planar one should also account for this deformation. Besides, in the case of objects with steep angles (where the incident collimated beam is far from normal to the metasurface at some points), further considerations should be taken in choosing the lattice constant to avoid excitation of higher order diffractions. Moreover, since the design procedure is local (i.e. each nano-post at each lattice site is chosen independently), the incident angle of the beam at each lattice point can be taken into account in designing the respective nano-post.

Conformal dielectric metasurfaces operate based on spatially varying nano-structured diffractive scatterers. The behavior of the device is wavelength dependent because both the optical response of the scatterers and their arrangement is optimized for a given wavelength. The performance of the proposed devices has a wavelength dependence similar to other high contrast transmitarray lenses recently demonstrated \cite{arbabi2015subwavelength} where good performance is maintained over a bandwidth of a few percent around the design wavelength.

The proposed platform is relatively robust to systematic and random errors. Fabrication errors do not affect the device functionality and only reduce its efficiency (5 nm error in nano-post diameters results in $\sim$3$\%$ reduction of efficiency \cite{arbabi2015dielectric}). Alignment imperfections (extra stretch or angular rotation) results in focal distance mismatch between the nonplanar object and the metasurface. Microlens focal distance has second order dependence on the substrate stretch ratio. For instance, for the devices shown in Fig. \ref{fig:Fig4_Measurement}, having 1$\%$ strain in the flexible metasurface results in a 2$\%$ error in focal distance and a 1 degree rotation misalignment results in 0.06$\%$ mismatch between the horizontal and vertical focal distances.   Also, fractional wavelength error is equal to the fractional error of the focal distance \cite{arbabi2015subwavelength} (i.e. 1$\%$ error in wavelength results in ~1$\%$ error in focal distance of the flexible metasurface).

The developed fabrication process has a near unity yield (more than 99.5$\%$) and we are able to transfer almost all of the nano-posts into the PDMS with good accuracy. Nevertheless, the proposed platform is very robust to the fabrication deficiencies; various imperfections including deviations between designed and fabricated nano-post sizes ($\sim$5 nm in the diameter and (or) height of the nano-posts), rough side walls, and missing nano-posts ($\sim$10$\%$) only result in small reductions in the efficiency of the device, and does not alter the functionality significantly.

In conclusion, we demonstrated flexible dielectric metasurfaces and showed their applications for conformal optics. As proof of concept, the optical properties of glass cylinders have been changed to behave like aspherical lenses focusing light to a point. The design paradigm can be applied to any other system where conformal optical design is required. In addition, flexible electronics is a well-established field of research, with the aim of transferring conventional systems to flexible and non-planar substrates. Very promising results have been achieved during the last decade with various applications in wearable electronics, electronic skins, and medical devices \cite {he2015inorganic,kim2008stretchable,wang2013user}.  The flexible and conformal metasurface platform proposed here can be merged with conformal electronics leading to versatile flexible optoelectronic technologies.

\clearpage
\section*{Methods}
\noindent\textbf{Design procedure}

The optical path length and the corresponding optical path difference of light passing through the cylinders were computed using ray optics approximation. For simulations, the convex and concave cylinders were assumed to have radii of 4.13 mm and 6.48 mm, respectively, and a refractive index of 1.507. The PDMS layer was modeled as a 50-$\mu$m-thick layer with a refractive index of 1.41. In both cases, the object OPDs were calculated at the outermost surface of the PDMS, considering light propagation through the PDMS layer and refraction at the glass-PDMS interface. The desired OPDs were also calculated at the same surfaces, assuming focal distances of 3.5 mm and 8 mm for the convex and concave lenses, respectively. Two different metasurfaces of diameters 1 mm and 2 mm were designed for the convex and concave cylinders to impose the phase shifts equivalent to the difference of the cylinders' and the desired OPDs. 

The planar periodic metasurfaces were simulated using the rigorous coupled wave analysis (RCWA) technique to find the complex transmission coefficients corresponding to all nano-post diameters for normal incident plane waves (Fig. \ref{fig:Fig2_Design}g) \cite{liu2012s}. The lattice constant is chosen such that the array is non-diffractive at the simulation wavelength. Simulation results shown in Supplementary Fig. 2 were also obtained using the RCWA technique. All of the simulations and calculations were performed at the wavelength of 915 nm. 

\noindent\textbf{Sample fabrication}\\
A 300-nm-thick germanium sacrificial layer was deposited by electron-beam evaporation on a silicon wafer, and 720 nm hydrogenated a-Si was deposited on the germanium layer using plasma enhanced chemical vapor deposition (PECVD) with a 5\% mixture of silane in argon at $200\,^{\circ}{\rm C}$. The refractive index of the a-Si layer was measured using variable angle spectroscopic ellipsometry and was found to be 3.56 at the wavelength of 915 nm. The metasurface pattern was defined in ZEP-520A positive resist ($\sim$300 nm, spin coated at 5000 rpm for 1 min) using a Vistec EBPG5000+ electron-beam lithography system. The pattern was developed in a resist developer (ZED-N50 from Zeon Chemicals). After developing the resist, the pattern was transferred into a $\sim$100-nm-thick aluminum oxide layer deposited by electron-beam evaporation through a lift-off process. The patterned aluminum oxide served as a hard mask for dry etching of the a-Si layer in a mixture of $\mathrm{SF_6}$ and $\mathrm{C_4F_8}$ plasma. The PDMS polymer (RTV-615 A and B mixed with a 10:1 mass ratio) was diluted in toluene in a 2:3 weight ratio as a thinner. The mixture was spin coated at 3000 rpm for 1 min on the fabricated metasurface to fill the gaps between the nano-posts and to form a thin PDMS film (Supplementary Fig. 3). The sample was degassed and cured for more than 30 mins. The second layer of PDMS without a thinner was spin coated on the sample to form a $\sim$50-$\mu$m-thick PDMS film (spin coated at 1000 rpm for 1 min). The sample was degassed and cured for more than 1 hr. Finally, immersion in a 1:1:30 mixture of ammonium hydroxide, hydrogen peroxide, and deionized water at room temperature removed the sacrificial germanium layer releasing the PDMS substrate and the embedded nano-posts ($\sim$one day). The released metasurface is then mounted manually on the cylinders (Edmund Optics 43-856 and 47-748). To compensate for the misalignment of the substrate and metasurface, multiple lenses with slightly different rotations were fabricated in each sample (Fig. \ref{fig:Fig3_Fab}c). This way, the best aligned microlens should have a rotation error of less than or equal to one degree (the rotation step between two successive metasurface lenses).

\noindent\textbf{Measurement procedure}\\
Devices were characterized using the setups shown schematically in Supplementary Fig. 7. A 915 nm fiber coupled semiconductor laser was used as the source and a fiber collimation package (Thorlabs F220APC-780) was used to collimate the beam. Intensity at different planes was captured by using a 50$\times$ objective lens (Olympus LMPlanFL N, NA=0.5), a tube lens (Thorlabs LB1945-B) with focal distance of 20 cm, and a camera (CoolSNAP K4 from Photometrics) as shown in Supplementary Fig. 7a. Moreover, neutral density (ND) filters (Thorlabs ND filters, B coated) were used to adjust the light intensity and decrease the background noise captured by the camera. The overall microscope magnification was measured by imaging a calibration sample with known feature sizes. To measure the efficiencies, an additional lens (Thorlabs LB1945-B with focal length of 20 cm) was used to partially focus the collimated beam, so that more than 99\% of the beam power falls inside the device under test. The beam radius was adjusted by changing the distance between the lens and the sample.  A 15 $\mu$m diameter pinhole (approximately three times the measured FWHM) was placed at the focal plane of the sample to only allow the light focused inside the pinhole area to pass through. The focusing efficiency was then determined as the ratio of measured optical power after the pinhole (i.e. the power in focus) to the measured power right before the sample (the incident power). The measurement setup used for efficiency characterization is shown in Supplementary Fig. 7b. For polarization sensitivity measurement, a polarizer (Thorlabs LPNIR050-MP) was added before the sample to set the polarization state of the incident beam.

The data that support the findings of this study are available from the corresponding author upon the request.

\section*{Acknowledgement}
This work was supported by the DOE ``Light-Material Interactions in Energy Conversion" Energy Frontier Research Center funded by the US Department of Energy, Office of Science, Office of Basic Energy Sciences under Award no. DE-SC0001293. A.A. and E.A. were supported by Samsung Electronics. A.A. and Y.H were also supported by DARPA. The device nanofabrication was performed at the Kavli Nanoscience Institute at Caltech.

\clearpage

\newcommand{\beginsupplement}{%
        \setcounter{table}{0}
        \renewcommand{\thetable}{S\arabic{table}}%
        \setcounter{figure}{0}
        \renewcommand{\thefigure}{S\arabic{figure}}%
     }

\onecolumngrid

      \beginsupplement

\section{Supplementary Information for ``Decoupling optical function and geometrical form using conformal flexible dielectric metasurfaces"}

\renewcommand{\figurename}{\textbf{Supplementary Figure}}

\begin{figure*}[htp]
\centering
\includegraphics[width=\columnwidth]{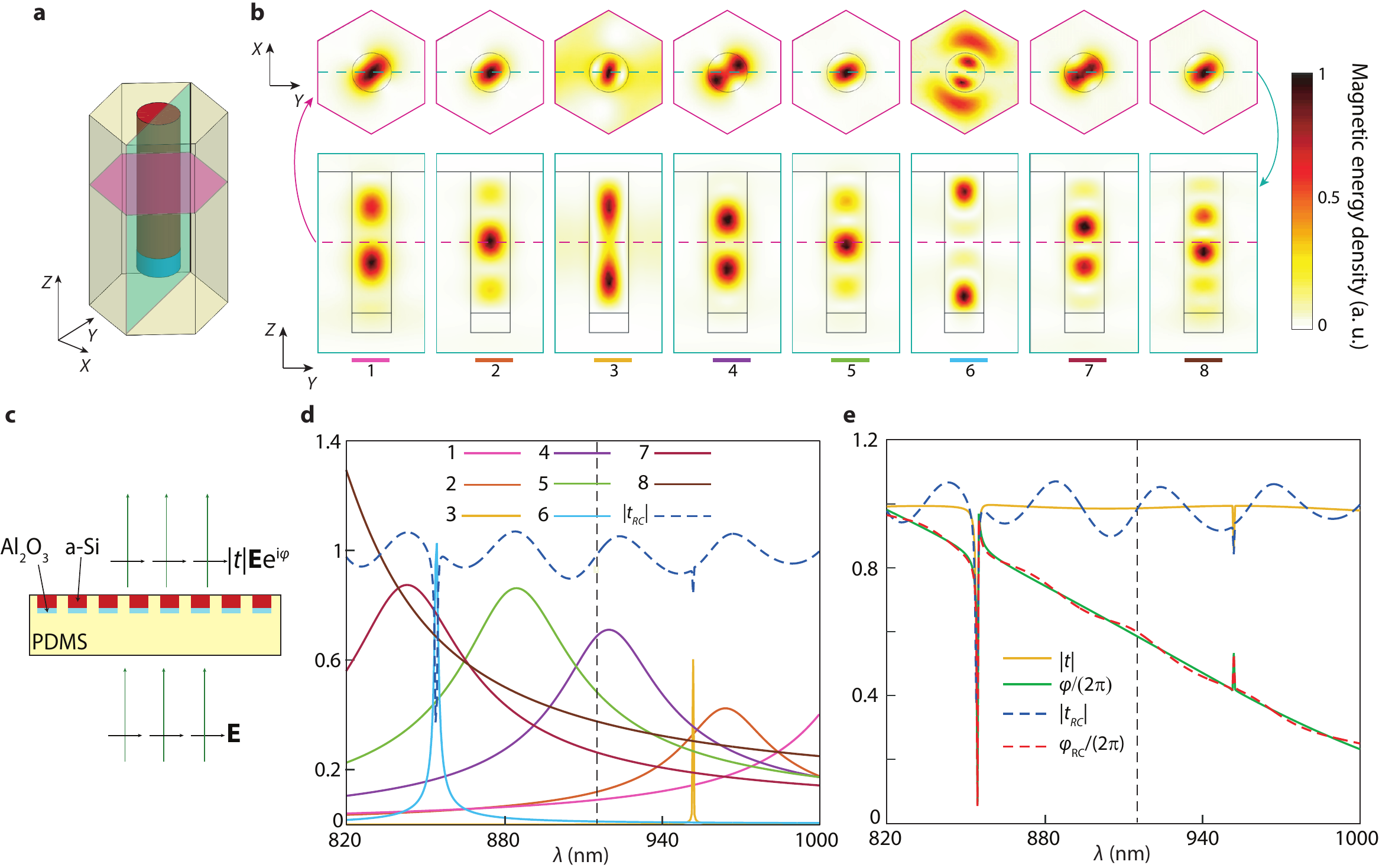}
\caption{\textbf{Resonant modes of the nano-posts and their contribution to transmission.} \textbf{(a)} Schematic illustration of the metasurface unit cell. \textbf{(b)} Magnetic energy density distribution of 8 dominant resonant modes in the bandwidth from 820 nm to 1000 nm, at horizontal (top) and vertical (bottom) cross sections shown in \textbf{a}. \textbf{(c)} Schematic illustration of a uniform array of nano-posts illuminated with a normally incident plane wave. The amplitude of the transmission coefficient ($\it{|t|}$) and its phase ($\it{\varphi}$) are indicated in the illustration. \textbf{(d)} Weighted resonance curves demonstrating contribution of 8 dominant resonant modes to the transmission of a periodic array of nano-posts with diameter of 200 nm, as well as the reconstructed transmission amplitude from these 8 modes. \textbf{(e)} Transmission amplitude and phase of the periodic array of nano-posts, and  reconstructed transmission amplitude and phase using the 8 dominant resonant modes. $|t_\mathrm{RC}|$: reconstructed transmission amplitude, $\varphi_\mathrm{RC}$: reconstructed transmission phase.}
\end{figure*}

\begin{figure*}[htp]
\centering
\includegraphics[width=0.9\columnwidth]{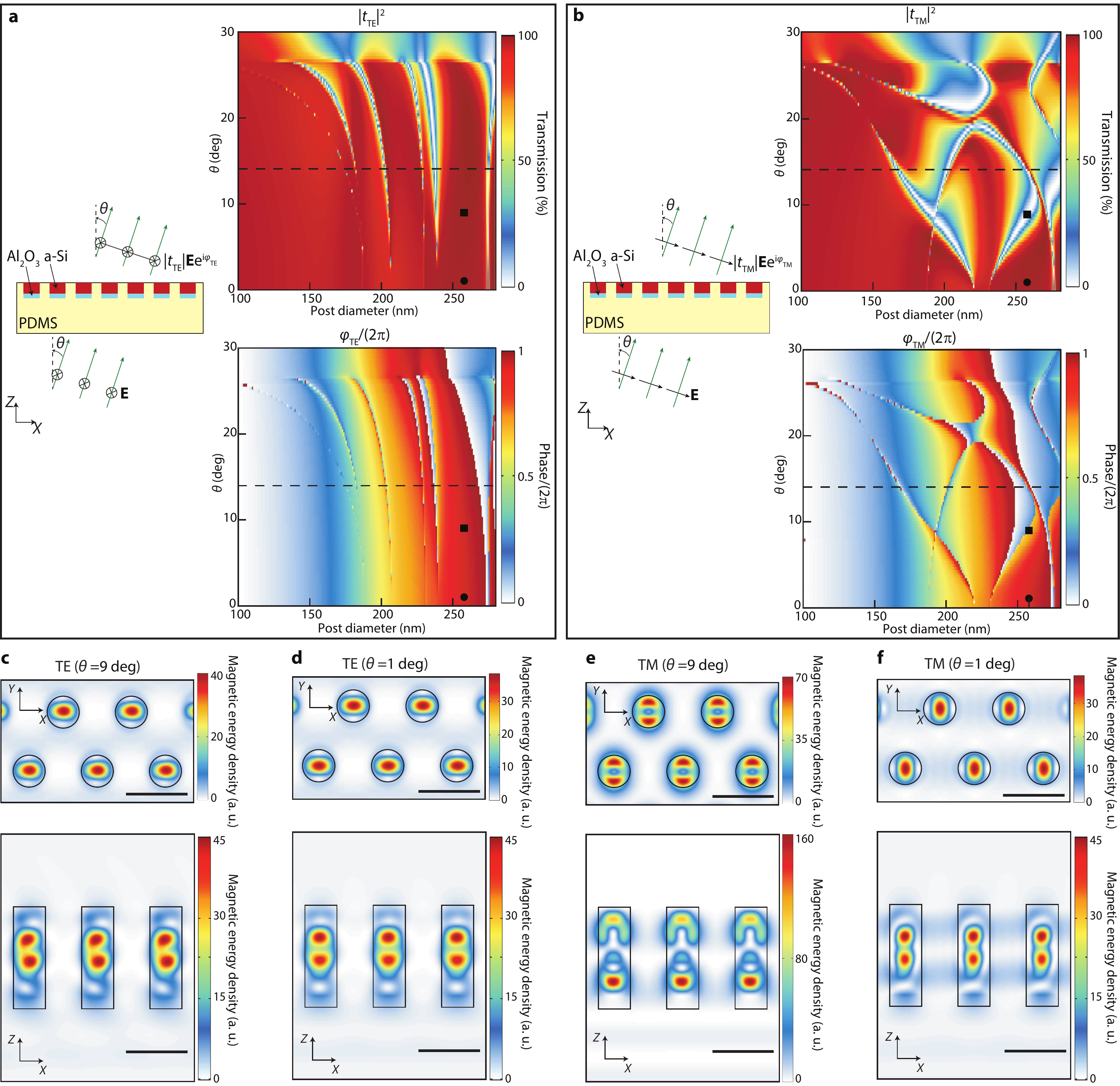}
\caption{\textbf{Angular dependence of the transmission coefficient.} \textbf{(a)} Schematic illustration of a uniform array of nano-posts embedded in polydimethylsiloxane (PDMS) illuminated by an obliquely incident plane wave (left), and its simulated transmission amplitude and phase as a function of nano-post diameter and incident beam angle for TE and TM (\textbf{b}) polarizations. Black dashed lines indicate the largest angle between the metasurface normal and the incident beam used in this manuscript. \textbf{(c)} Top and side views of magnetic energy density distribution for TE polarization under an oblique incident plane wave with $\it{\theta}$=9$^{\circ}$(solid squares in \textbf{a}) and \textbf{(d)} $\it{\theta}$=1$^{\circ}$(solid circles in \textbf{a}). \textbf{(e)} Top and side views of magnetic energy density distribution for TM polarization under an incident plane wave with $\it{\theta}$=9$^{\circ}$ (solid squares in \textbf{b}) and \textbf{(f)} $\it{\theta}$=1$^{\circ}$ (solid circles in \textbf{b}). Black solid lines show the boundaries of the nano-posts. The magnetic energy density of the incident plane wave is chosen to be 1. Scale bars, 500 nm. TE: transverse electric, TM: transverse magnetic.}
\end{figure*}

\begin{figure*}[htp]
\centering
\includegraphics[width=\columnwidth]{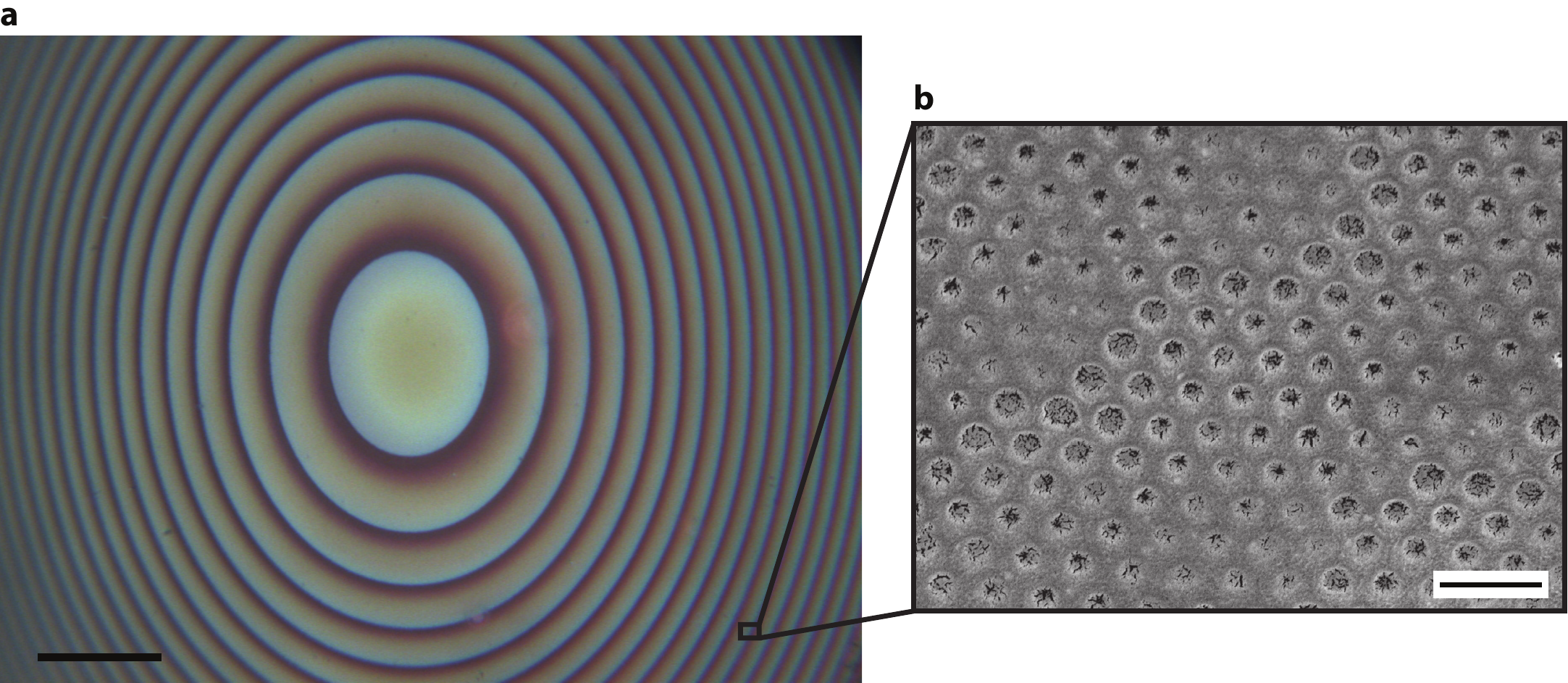}
\caption{\textbf{Complete embedding of amorphous silicon nano-posts in PDMS.} \textbf{(a)} Optical microscope image of a portion of a fabricated conformal metasurface lens after transferring to a flexible substrate. This image shows that almost all of the nano-posts are retained in the transfer process. Scale bar, 100 $\mu$m. \textbf{(b)} Scanning electron microscope image of a portion of the flexible metasurface, taken at a tilt angle of 30 degrees. The image shows silicon nano-posts are entirely embedded in the flexible substrate (polydimethylsiloxane (PDMS)), and void-free filling of the gaps between the nano-posts with PDMS. To dissipate charge accumulation during scanning electron imaging, a $\sim$15-nm-thick gold layer was deposited on the sample prior to imaging. The cracks seen in the gold layer at the position of the nano-posts were not present initially and were gradually formed as the sample was exposed to the electron beam. Scale bar, 1 $\mu$m.}
\end{figure*}

\begin{figure*}[htp]
\centering
\includegraphics[width=\columnwidth]{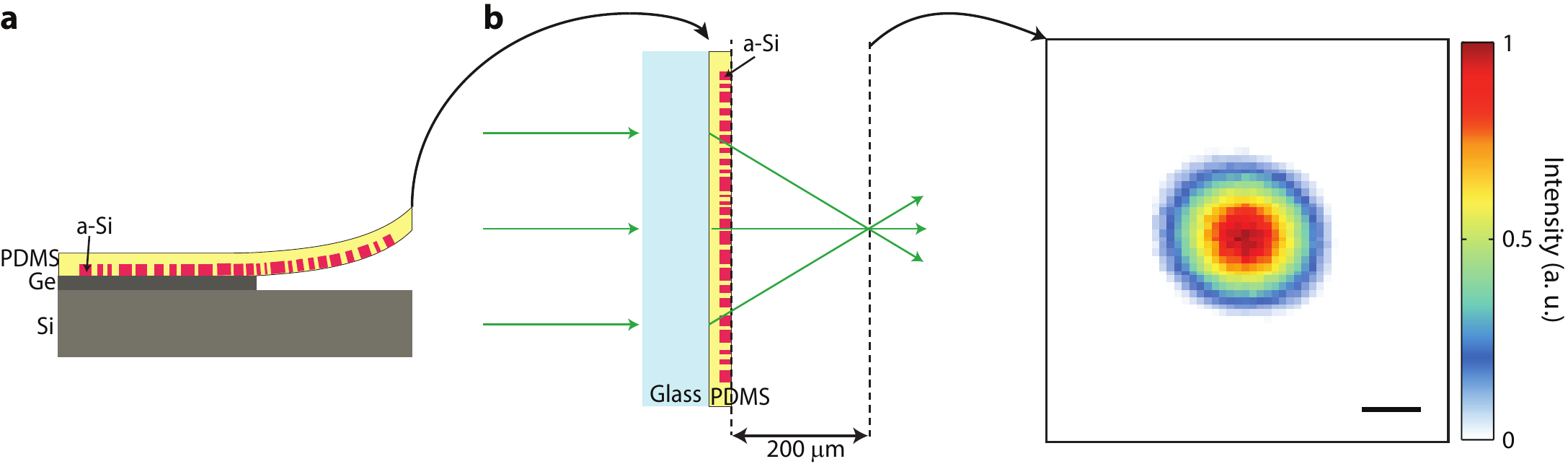}
\caption{\textbf{Preservation of high efficiency and diffraction limited optical performance of the metasurfaces through the transfer process.} \textbf{(a)} A metasurface lens designed to operate as an aspherical lens when mounted on a flat substrate is transferred to a polydimethylsiloxane (PDMS) substrate. The metasurface lens has a diameter of 200 $\mu$m and a focal distance of 200 $\mu$m. \textbf{(b)} Measured focal plane intensity profile when the flexible metasurface is mounted on a flat glass substrate and illuminated with a collimated beam (as shown in the inset). The measured full width at half maximum (FWHM) spot size of $\sim$ 1 $\mu$m agrees well with the diffraction limited FWHM spot size of 1 $\mu$m. The focusing efficiency of the lens was measured as 70\%. The diffraction limited spot size and the relatively high focusing efficiency verifies the fidelity of the fabrication process in preserving optical properties of metasurfaces. Measurements are performed at the wavelength of 915 nm (see Methods for measurement details). Scale bar, 1 $\mu$m.}
\end{figure*}

\begin{figure*}[htp]
\centering
\includegraphics[width=\columnwidth]{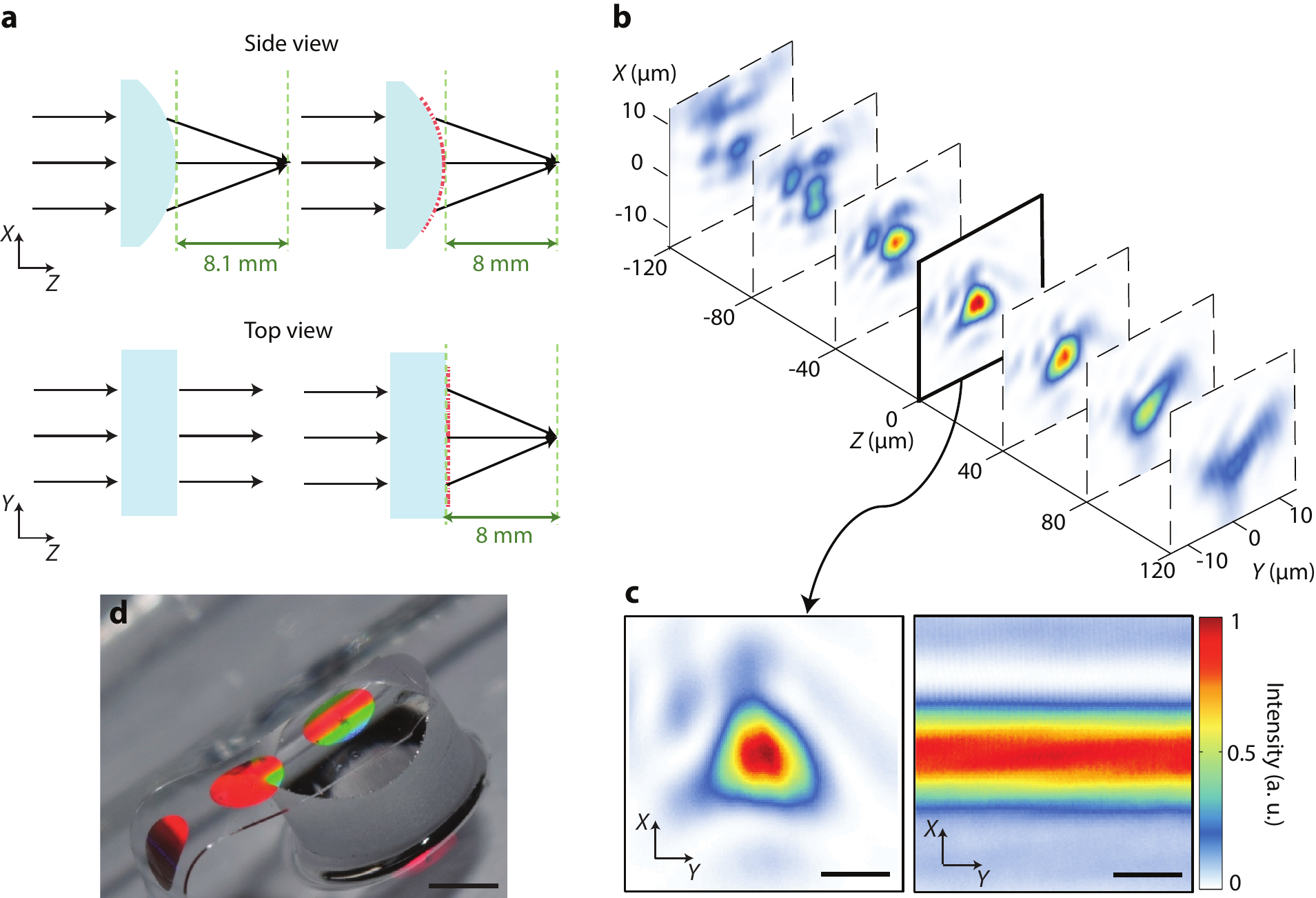}
\caption{\textbf{Conformal metasurface with steep incident angles.} \textbf{(a)} Schematic illustration of a metasurface which converts a cylindrical lens to an aspheric lens. The metasurface diameter is 2 mm, and the largest angle between the metasurface normal and the incident beam is 14$^{\circ}$. \textbf{(b)} Intensities measured at different planes parallel to the focal plane of the cylinder metasurface combination. \textbf{(c)} Measured intensities for the glass cylinder (right) and cylinder plus metasurface combination (left) at their respective focal planes. The measured full width at half maximum (FWHM) spot size at the focal plane is approximately 4.5 $\mu$m, which is comparable to the diffraction limited FWHM spot size of 3.7 $\mu$m. For the metasurface cylinder combination,
by using a setup shown in Supplementary Fig. 7b, focusing efficiencies of 68\% and 64\% were measured for transverse electric (TE) and transverse magnetic (TM) polarizations, respectively. All the measurements are performed at the wavelength of 915 nm. Scale bars, 5 $\mu$m. \textbf{(d)} Optical image of the fabricated metasurface mounted on the glass cylinder. Scale bar, 2 mm.}
\end{figure*}

\begin{figure*}[htp]
\centering
\includegraphics[width=0.65\columnwidth]{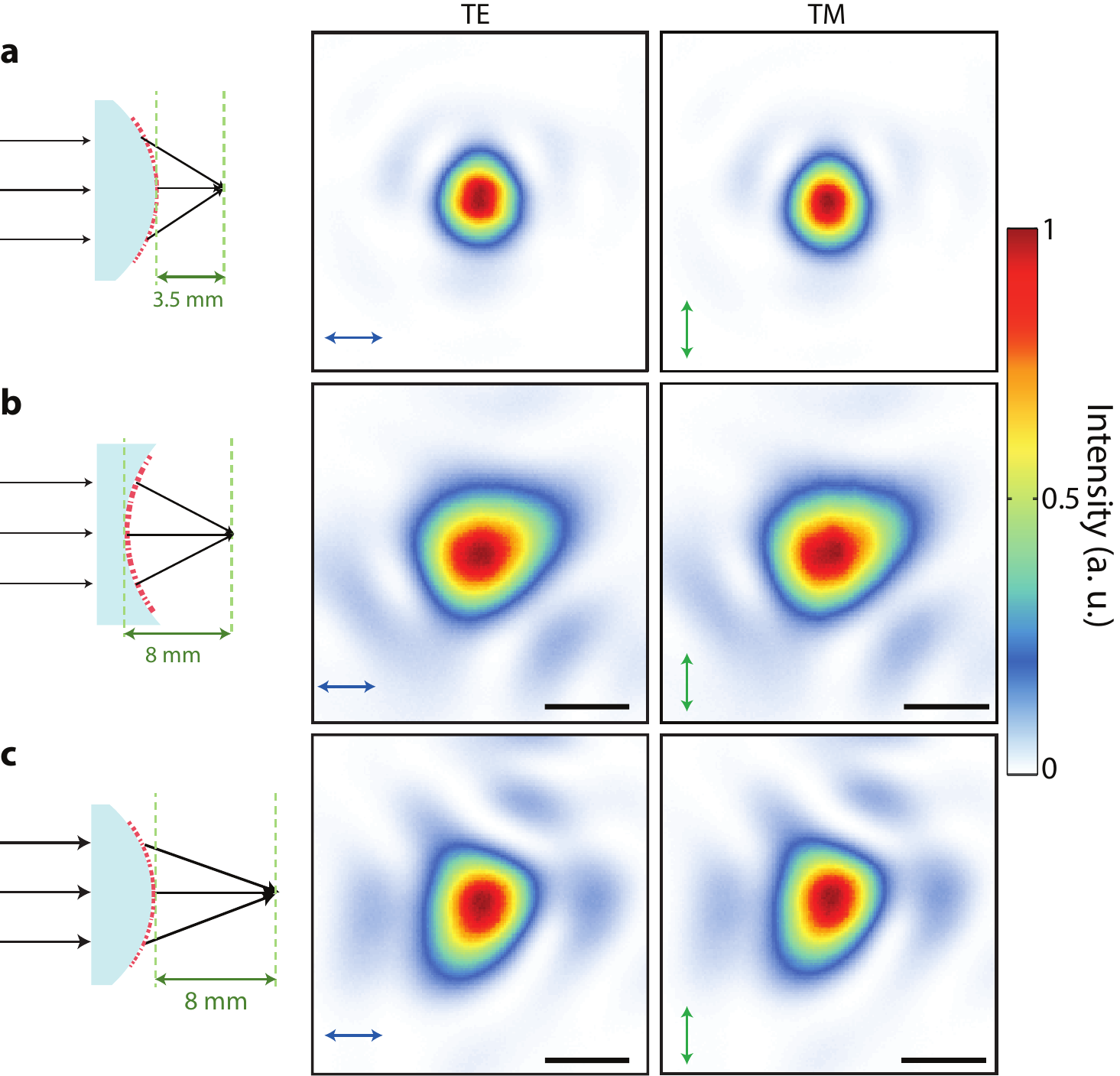}
\caption{\textbf{Effect of the input beam polarization on device performance.} Intensity pattern measured at the focal plane of three different conformal metasurfaces with TE (left) and TM (right) polarizations. \textbf{(a)} A 1 mm-diameter metasurface on a convex cylinder with a 3.5 mm focal distance, \textbf{(b)} A 2-mm-diameter metasurface on a concave cylinder with an 8 mm focal distance, and \textbf{(c)} A 2-mm-diameter metasurface on a convex cylinder with an 8 mm focal distance. Focus pattern shows very negligible polarization dependence for all of the devices. The measured efficiencies, however, are more sensitive to polarization. The 1-mm-diameter metasurface with a maximum beam incident angle of 7$^{\circ}$ has an efficiency of 56\% for both polarizations, while for the 2-mm-diameter metasurface lens on convex cylinder with maximum incident angle of 14$^{\circ}$ the efficiency drops from 68\% for TE polarization to 64\% for TM polarization. Besides, the metasurface lens on the concave cylinder with a maximum beam incident angle of 9$^{\circ}$ has efficiencies of 52\% and 50\% for TE and TM polarizations,  respectively. This is in accordance with the angular dependence of transmission coefficient as shown in Supplementary Fig. 2. The device efficiency for TM polarization degrades as the angle between metasurface normal and incident beam increases. However, this does not considerably affect the focus shape for
any of the devices. Scale bars, 5 $\mu$m. TE: transverse electric, TM: transverse magnetic.}
\end{figure*}

\begin{figure*}[htp]
\centering
\includegraphics[width=\columnwidth]{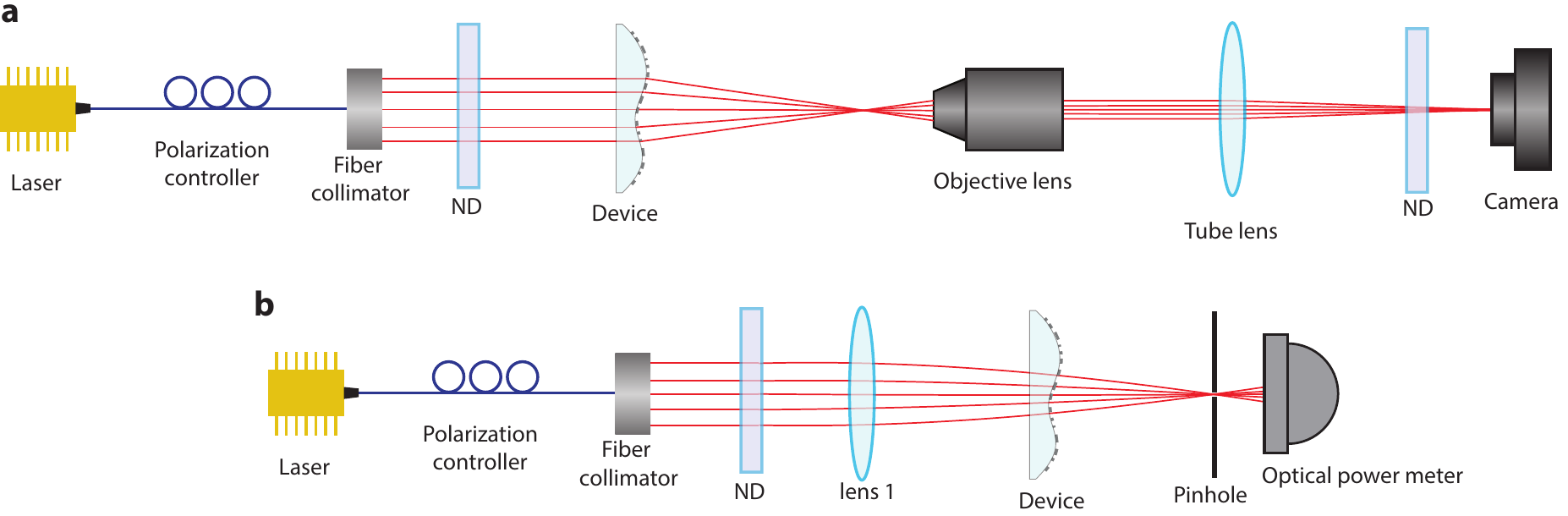}
\caption{\textbf{Measurement setup}. \textbf{(a)} Schematic illustration of the measurement setup used for characterization of the conformal metasurfaces. ND: neutral density filter. \textbf{(b)} Schematic diagram of the experimental setup used for measuring the efficiencies of conformal metasurfaces.}
\end{figure*}

\clearpage
\section{Supplementary Note 1: Resonant modes of the nano-posts}
To get more insight into the resonant modes contributing to the transmission coefficient, we consider the amplitude and phase of the transmission coefficient of a periodic array of the nano-posts as a function of wavelength around 915 nm (yellow and green solid curves in Supplementary Fig. 1e). The nano-posts diameter was chosen to be 200 nm, in the middle of our parameter scale. The resonant modes of such an array that have resonant frequencies within the desired bandwidth are found using the finite element method in COMSOL Multiphysics. The magnetic energy density for these resonant modes is shown in supplementary Fig. 1b in two horizontal and vertical cross sections indicated in supplementary Fig. 1a. Total transmission of the nano-posts array is determined by the interference between the incident light and these dominant resonant modes. We have reconstructed the amplitude and phase of the transmission coefficient in the bandwidth of interest using these resonant curves. Supplementary Fig. 1d shows the resonant curves with their corresponding amplitudes used in the reconstruction, along with the reconstructed transmission amplitude. Supplementary Fig. 1e compares the actual transmission amplitude and phase of the nano-posts array with the reconstructed ones. Using more resonant modes (with frequencies outside the bandwidth of interest) it is possible to reconstruct the transmission with even higher accuracy.
\section{Supplementary Note 2: Angular dependence of transmission coefficient for TE and TM polarizations}
Supplementary Fig. 2a shows a schematic illustration of a uniform array of nano-posts embedded in PDMS (left), and its simulated transmission amplitude (top) and phase (bottom) as a function of nano-posts diameter and incident beam angle for TE polarization. Similar plots for TM polarization are shown in Supplementary Fig. 2b. These results show weak angular dependence of the high contrast transmitarray metasurfaces for TE polarization in the range of angles involved in their operation. 
The larger angular dependence for the TM polarization results in slight degradation of the device performance in this polarization. The maximum angle between the metasurface normal and the incident beam for the two metasurfaces discussed in the main text are 7$^{\circ}$ and 9$^{\circ}$. The angular dependence increases for larger incident angles, indicating that angular dependence should be considered in the design of conformal metasurfaces with steep local incident angles. Moreover, for the lattice used here, higher diffraction orders are present for incident angles larger than $\sim$25$^{\circ}$, and a smaller lattice constant should be used for metasurfaces with larger incident angles.
Magnetic energy density distributions for a uniform nano-posts array with the diameter of 258 nm illuminated with a plane wave with incident angle of $\it{\theta}=$9$^{\circ}$ for TE and TM polarizations are shown in Supplementary Fig. 2c and 2e, respectively. The corresponding diameter/angle points are indicated by solid squares on the transmission plots presented in Supplementary Figs. 2a and 2b. Similar plots of the magnetic energy density distributions for an incident angle $\it{\theta}$=1$^{\circ}$ are shown in Supplementary Figs. 2d and 2f (indicated by solid circles on the transmission plots in Supplementary Figs. 2a and 2b). The magnetic energy density of the incident beam is chosen to be 1. For obliquely incident beams the symmetry between the TE and TM waves is broken, and this results in various modes in the nano-posts being excited with different amplitudes for the two polarizations. This in turn results in dissimilar behavior for the TE and TM polarizations as a function of the incident angle. Comparing Supplementary Fig. 2c and 2d, we can observe that the magnetic energy density inside the nano-posts is very similar for TE wave with different incident angles. From supplementary Fig. 2e and 2f, however, we can see that the magnetic energy density distribution changes significantly with incident angle for TM wave. This change in excited modes of the nano-posts for the TM waves as a function of angle, which is a consequence of the axial component of the electric field becoming important, results in the stronger angular dependence of the TM transmission observed here.
\section{Supplementary Note 3: High fidelity of the nano-post transfer process}
To preserve the high efficiency and the diffraction limited optical performance of the metasurfaces through the transfer process to the PDMS substrate, it is essential that a large majority of the nano-posts are transferred, and the gaps between the nano-posts are uniformly filled with PDMS. The efficiency of the metasurface decreases and wavefront aberrations are introduced if some of the nano-posts are not transferred because the portion of the light passing through the areas with missing nano-posts does not undergo the proper phase shift. Air voids between the nano-posts also degrade the efficiency and cause wavefront aberrations by disturbing the near-field optical distributions of the nano-posts which in turn leads to a lower local transmission efficiency and an incorrect phase shift. To verify that all the nano-posts are transferred to the flexible substrate, after the transfer process, the metasurfaces were examined using an optical microscope. We observed that all the nano-posts were successfully transferred as it is shown in Supplementary Fig. 3a, which shows an optical microscope image of a portion of a flexible metasurface. The void-free filling of the gaps between the nano-posts with PDMS was verified using scanning electron microscopy (SEM). A thin layer of gold ($\sim$15 nm) was deposited on the top surface of the flexible metasurface to avoid charge accumulation during SEM imaging. An SEM image of the metasurface taken at a tilt angle of 30 degrees with respect to the metasurface normal is shown in Supplementary Fig. 3b. The cracks seen in this SEM image at the location of the nano-posts are in the gold layer. They did not initially exist and were slowly appeared as the imaged area was exposed to the electron beam and the sample was charged. As Supplementary Fig. 3b shows, no void is present between the nano-posts, and the same conclusion was made when a larger area of the metasurface was inspected with SEM. According to all different SEM images, the yield of the entire fabrication process is more than 99.5$\%$ with a 95$\%$ confidence interval \cite{sauro2005estimating}.

\section{Supplementary References}


\clearpage
\section*{References}
\begin{thebibliography}{10}
\expandafter\ifx\csname url\endcsname\relax
  \def\url#1{\texttt{#1}}\fi
\expandafter\ifx\csname urlprefix\endcsname\relax\def\urlprefix{URL }\fi
\providecommand{\bibinfo}[2]{#2}
\providecommand{\eprint}[2][]{\url{#2}}

\bibitem{born1999principles}
\bibinfo{author}{Born, M.} \& \bibinfo{author}{Wolf, E.}
\newblock \emph{\bibinfo{title}{{Principles of Optics}}}
  (\bibinfo{publisher}{Cambridge Univ. Press},
  \bibinfo{year}{1999}).

\bibitem{thompson2012freeform}
\bibinfo{author}{Thompson, K.~P.} \& \bibinfo{author}{Rolland, J.~P.}
\newblock \bibinfo{title}{Freeform optical surfaces: a revolution in imaging
  optical design}.
\newblock \emph{\bibinfo{journal}{Opt. Photon. News}}
  \textbf{\bibinfo{volume}{23}}, \bibinfo{pages}{30--35}
  (\bibinfo{year}{2012}).
\bibitem{shannonSPIE1999}
\bibinfo{author}{Shannon, R.~R.}
\newblock \bibinfo{title}{Overview of conformal optics}.
\newblock \emph{\bibinfo{journal}{Proc. SPIE}}
  \textbf{\bibinfo{volume}{3705}}, \bibinfo{pages}{180--188}
  (\bibinfo{year}{1999}).  
\bibitem{KanppSPIE2002}
\bibinfo{author}{Knapp, D.~J.}
\newblock \bibinfo{title}{Fundamentals of conformal dome design}.
\newblock \emph{\bibinfo{journal}{Proc. SPIE}}
  \textbf{\bibinfo{volume}{4832}}, \bibinfo{pages}{394--409}
  (\bibinfo{year}{2002}).
  
\bibitem{ni2015ultrathin}
\bibinfo{author}{Ni, X.}, \bibinfo{author}{Wong, Z.~J.},
  \bibinfo{author}{Mrejen, M.}, \bibinfo{author}{Wang, Y.} \&
  \bibinfo{author}{Zhang, X.}
\newblock \bibinfo{title}{An ultrathin invisibility skin cloak for visible
  light}.
\newblock \emph{\bibinfo{journal}{Science}} \textbf{\bibinfo{volume}{349}},
  \bibinfo{pages}{1310--1314} (\bibinfo{year}{2015}).
  
  \bibitem{fan2012invisible}
\bibinfo{author}{Fan, P.} \emph{et~al.}
\newblock \bibinfo{title}{An invisible metal-semiconductor photodetector}.
\newblock \emph{\bibinfo{journal}{Nat. Photon.}}
  \textbf{\bibinfo{volume}{6}}, \bibinfo{pages}{380--385}
  (\bibinfo{year}{2012}).
  
\bibitem{valentine2009optical}
\bibinfo{author}{Valentine, J.}, \bibinfo{author}{Li, J.},
  \bibinfo{author}{Zentgraf, T.}, \bibinfo{author}{Bartal, G.} \&
  \bibinfo{author}{Zhang, X.}
\newblock \bibinfo{title}{An optical cloak made of dielectrics}.
\newblock \emph{\bibinfo{journal}{Nat. Mater.}}
  \textbf{\bibinfo{volume}{8}}, \bibinfo{pages}{568--571}
  (\bibinfo{year}{2009}).

\bibitem{ergin2010three}
\bibinfo{author}{Ergin, T.}, \bibinfo{author}{Stenger, N.},
  \bibinfo{author}{Brenner, P.}, \bibinfo{author}{Pendry, J.~B.} \&
  \bibinfo{author}{Wegener, M.}
\newblock \bibinfo{title}{Three-dimensional invisibility cloak at optical
  wavelengths}.
\newblock \emph{\bibinfo{journal}{Science}} \textbf{\bibinfo{volume}{328}},
  \bibinfo{pages}{337--339} (\bibinfo{year}{2010}).
  

\bibitem{teo2015controlling}
\bibinfo{author}{Teo, J. Y.~H.}, \bibinfo{author}{Wong, L.~J.},
  \bibinfo{author}{Molardi, C.} \& \bibinfo{author}{Genevet, P.}
\newblock \bibinfo{title}{Controlling electromagnetic fields at boundaries of
  arbitrary geometries}.
\newblock Preprint at http://arxiv.org/abs/{1509.06175} (\bibinfo{year}{2015}).  
  \bibitem{zheng2015metasurface}
  
  
\bibinfo{author}{Zheng, G.} \emph{et~al.}
\newblock \bibinfo{title}{Metasurface holograms reaching 80\% efficiency}.
\newblock \emph{\bibinfo{journal}{Nat. Nanotech.}}
  \textbf{\bibinfo{volume}{10}}, \bibinfo{pages}{308--312}
  (\bibinfo{year}{2015}).

\bibitem{kildishev2013planar}
\bibinfo{author}{Kildishev, A.~V.}, \bibinfo{author}{Boltasseva, A.} \&
  \bibinfo{author}{Shalaev, V.~M.}
\newblock \bibinfo{title}{Planar photonics with metasurfaces}.
\newblock \emph{\bibinfo{journal}{Science}} \textbf{\bibinfo{volume}{339}},
  \bibinfo{pages}{1232009} (\bibinfo{year}{2013}).

\bibitem{yu2014flat}
\bibinfo{author}{Yu, N.} \& \bibinfo{author}{Capasso, F.}
\newblock \bibinfo{title}{Flat optics with designer metasurfaces}.
\newblock \emph{\bibinfo{journal}{Nat. Mater.}}
  \textbf{\bibinfo{volume}{13}}, \bibinfo{pages}{139--150}
  (\bibinfo{year}{2014}).

\bibitem{Lin2014a}
\bibinfo{author}{Lin, D.}, \bibinfo{author}{Fan, P.}, \bibinfo{author}{Hasman,
  E.} \& \bibinfo{author}{Brongersma, M.~L.}
\newblock \bibinfo{title}{{Dielectric gradient metasurface optical elements}}.
\newblock \emph{\bibinfo{journal}{Science}} \textbf{\bibinfo{volume}{345}},
  \bibinfo{pages}{298--302} (\bibinfo{year}{2014}).

\bibitem{arbabi2015dielectric}
\bibinfo{author}{Arbabi, A.}, \bibinfo{author}{Horie, Y.},
  \bibinfo{author}{Bagheri, M.} \& \bibinfo{author}{Faraon, A.}
\newblock \bibinfo{title}{Dielectric metasurfaces for complete control of phase
  and polarization with subwavelength spatial resolution and high
  transmission}.
\newblock \emph{\bibinfo{journal}{Nat. Nanotech.}}
\textbf{\bibinfo{volume}{10}},
  \bibinfo{pages}{937-–943} (\bibinfo{year}{2015}).


\bibitem{karimi2014generating}
\bibinfo{author}{Karimi, E.} \emph{et~al.}
\newblock \bibinfo{title}{Generating optical orbital angular momentum at
  visible wavelengths using a plasmonic metasurface}.
\newblock \emph{\bibinfo{journal}{Light Sci. Appl.}}
  \textbf{\bibinfo{volume}{3}}, \bibinfo{pages}{e167} (\bibinfo{year}{2014}).

\bibitem{Fattal2010}
\bibinfo{author}{Fattal, D.}, \bibinfo{author}{Li, J.}, \bibinfo{author}{Peng,
  Z.}, \bibinfo{author}{Fiorentino, M.} \& \bibinfo{author}{Beausoleil, R.~G.}
\newblock \bibinfo{title}{{Flat dielectric grating reflectors with focusing
  abilities}}.
\newblock \emph{\bibinfo{journal}{Nat. Photon.}}
  \textbf{\bibinfo{volume}{4}}, \bibinfo{pages}{466--470}
  (\bibinfo{year}{2010}).

\bibitem{arbabi2015subwavelength}
\bibinfo{author}{Arbabi, A.}, \bibinfo{author}{Horie, Y.},
  \bibinfo{author}{Ball, A.~J.}, \bibinfo{author}{Bagheri, M.} \&
  \bibinfo{author}{Faraon, A.}
\newblock \bibinfo{title}{Subwavelength-thick lenses with high numerical
  apertures and large efficiency based on high-contrast transmitarrays}.
\newblock \emph{\bibinfo{journal}{Nat. Commun.}}
  \textbf{\bibinfo{volume}{6}}, \bibinfo{pages}{7069} (\bibinfo{year}{2015}).

\bibitem{Vo2014}
\bibinfo{author}{Vo, S.} \emph{et~al.}
\newblock \bibinfo{title}{{Sub-wavelength grating lenses with a twist}}.
\newblock \emph{\bibinfo{journal}{IEEE Photon. Technol. Lett.}}
  \textbf{\bibinfo{volume}{26}}, \bibinfo{pages}{1375--1378}
  (\bibinfo{year}{2014}).

\bibitem{di2010flexible}
\bibinfo{author}{Falco, A. D.}, \bibinfo{author}{Ploschner, M.} \&
  \bibinfo{author}{Krauss, T.~F.}
\newblock \bibinfo{title}{Flexible metamaterials at visible wavelengths}.
\newblock \emph{\bibinfo{journal}{New J. Phys.}}
  \textbf{\bibinfo{volume}{12}}, \bibinfo{pages}{113006}
  (\bibinfo{year}{2010}).

\bibitem{pryce2010highly}
\bibinfo{author}{Pryce, I. M.}, \bibinfo{author}{Aydin, K.},
  \bibinfo{author}{Kelaita, Y. A.}, \bibinfo{author}{Briggs, R. M.} \&
  \bibinfo{author}{Atwater, H. A.}
\newblock \bibinfo{title}{Highly strained compliant optical metamaterials with
  large frequency tunability}.
\newblock \emph{\bibinfo{journal}{Nano Lett.}} \textbf{\bibinfo{volume}{10}},
  \bibinfo{pages}{4222--4227} (\bibinfo{year}{2010}).

\bibitem{xu2011flexible}
\bibinfo{author}{Xu, X.} \emph{et~al.}
\newblock \bibinfo{title}{Flexible visible-infrared metamaterials and their
  applications in highly sensitive chemical and biological sensing}.
\newblock \emph{\bibinfo{journal}{Nano Lett.}} \textbf{\bibinfo{volume}{11}},
  \bibinfo{pages}{3232--3238} (\bibinfo{year}{2011}).

\bibitem{walia2015flexible}
\bibinfo{author}{Walia, S.} \emph{et~al.}
\newblock \bibinfo{title}{Flexible metasurfaces and metamaterials: A review of
  materials and fabrication processes at micro-and nano-scales}.
\newblock \emph{\bibinfo{journal}{Appl. Phys. Rev.}}
  \textbf{\bibinfo{volume}{2}}, \bibinfo{pages}{011303} (\bibinfo{year}{2015}).

\bibitem{zhu2015flexible}
\bibinfo{author}{Zhu, L.}, \bibinfo{author}{Kapraun, J.},
  \bibinfo{author}{Ferrara, J.} \& \bibinfo{author}{Chang-Hasnain, C.~J.}
\newblock \bibinfo{title}{Flexible photonic metastructures for tunable
  coloration}.
\newblock \emph{\bibinfo{journal}{Optica}} \textbf{\bibinfo{volume}{2}},
  \bibinfo{pages}{255--258} (\bibinfo{year}{2015}).

\bibitem{gao2015optics}
\bibinfo{author}{Gao, L.} \emph{et~al.}
\newblock \bibinfo{title}{Optics and nonlinear buckling mechanics in large-area, highly stretchable arrays of plasmonic nanostructures}.
\newblock \emph{\bibinfo{journal}{ACS Nano.}} \textbf{\bibinfo{volume}{9}},
  \bibinfo{pages}{5968--5975} (\bibinfo{year}{2015}).
  
\bibitem{gutruf2015mechanically}
\bibinfo{author}{Gutruf, P.} \emph{et~al.}
\newblock \bibinfo{title}{Mechanically tunable dielectric resonator metasurfaces at visible frequencies}.
\newblock \emph{\bibinfo{journal}{ACS Nano.}} 
\textbf{\bibinfo{volume}{10}},
  \bibinfo{pages}{133--141} (\bibinfo{year}{2015}).

\bibitem{josefsson2006conformal}
\bibinfo{author}{Josefsson, L.} \& \bibinfo{author}{Persson, P.}
\newblock \emph{\bibinfo{title}{Conformal Array Antenna Theory and Design}.}
  \textbf{\bibinfo{volume}{29}}, (\bibinfo{publisher}{John Wiley \& Sons},
  \bibinfo{year}{2006}).
  
\bibitem{lalanne1999waveguiding}
\bibinfo{author}{Lalanne, P.}
\newblock \bibinfo{title}{Waveguiding in blazed-binary diffractive elements}.
\newblock \emph{\bibinfo{journal}{JOSA A}}
  \textbf{\bibinfo{volume}{16}}, \bibinfo{pages}{2517--2520}
  (\bibinfo{year}{1999}).


\bibitem{decker2015high}
\bibinfo{author}{Decker, M.} \emph{et~al.}
\newblock \bibinfo{title}{High-efficiency dielectric Huygens' surfaces}.
\newblock \emph{\bibinfo{journal}{Adv. Opt. Mater.}} 
 \textbf{\bibinfo{volume}{3}}, \bibinfo{pages}{813--820}
(\bibinfo{year}{2015}).


\bibitem{lalanne1998blazed}
\bibinfo{author}{Lalanne, P.}, \bibinfo{author}{Astilean, S.}, \bibinfo{author}{Cambril, E.} \& \bibinfo{author}{Launois, H.}
\newblock \bibinfo{title}{Blazed binary subwavelength gratings with efficiencies larger than those of conventional {\'e}chelette gratings}.
\newblock \emph{\bibinfo{journal}{Opt. Lett.}}
  \textbf{\bibinfo{volume}{23}}, \bibinfo{pages}{1081--1083}
  (\bibinfo{year}{1998}).


\bibitem{he2015inorganic}
\bibinfo{author}{He, J.}, \bibinfo{author}{Nuzzo, R.~G.},
  \bibinfo{author}{Rogers, J.}
\newblock \bibinfo{title}{Inorganic materials and assembly techniques for
  flexible and stretchable electronics}.
\newblock \emph{\bibinfo{journal}{Proc. IEEE}}
  \textbf{\bibinfo{volume}{103}}, \bibinfo{pages}{619--632}
  (\bibinfo{year}{2015}).

\bibitem{kim2008stretchable}
\bibinfo{author}{Kim, D.} \emph{et~al.}
\newblock \bibinfo{title}{Stretchable and foldable silicon integrated circuits}.
\newblock \emph{\bibinfo{journal}{Science}}
  \textbf{\bibinfo{volume}{320}}, \bibinfo{pages}{507--511}
  (\bibinfo{year}{2008}).



\bibitem{wang2013user}
\bibinfo{author}{Wang, C.} \emph{et~al.}
\newblock \bibinfo{title}{User-interactive electronic skin for instantaneous
  pressure visualization}.
\newblock \emph{\bibinfo{journal}{Nat. Mater.}}
  \textbf{\bibinfo{volume}{12}}, \bibinfo{pages}{899--904}
  (\bibinfo{year}{2013}).

\bibitem{liu2012s}
\bibinfo{author}{Liu, V.} \& \bibinfo{author}{Fan, S.}
\newblock \bibinfo{title}{S4: A free electromagnetic solver for layered
  periodic structures}.
\newblock \emph{\bibinfo{journal}{Comput. Phys. Commun.}}
  \textbf{\bibinfo{volume}{183}}, \bibinfo{pages}{2233--2244}
  (\bibinfo{year}{2012}).

\end{thebibliography}

\begin{thebibliography}{10}
\expandafter\ifx\csname url\endcsname\relax
  \def\url#1{\texttt{#1}}\fi
\expandafter\ifx\csname urlprefix\endcsname\relax\def\urlprefix{URL }\fi
\providecommand{\bibinfo}[2]{#2}
\providecommand{\eprint}[2][]{\url{#2}}
\bibitem{sauro2005estimating}
\bibinfo{author}{Sauro, J.} \& \bibinfo{author}{Lewis, J. R.},
\newblock \bibinfo{title}{Estimating completion rates from small samples using binomial confidence intervals: comparisons and recommendations.}
\newblock \emph{\bibinfo{journal}{SAGE Publications}}
  \textbf{\bibinfo{volume}{49}}, \bibinfo{pages}{2100--2103}
  (\bibinfo{year}{2005}).
  \end{thebibliography}
\end{document}